# High-pressure phase transition of AB3-type compounds: case of tellurium trioxide

Dominik Kurzydłowski,[1] Mikhail A. Kuzovnikov,[2] Marek Tkacz[3]

[1] *Faculty of Mathematics and Natural Sciences, Cardinal Stefan Wyszyński University, Warsaw 01-038, Poland;*

[2] *Institute of Solid State Physics RAS, 142432 Chernogolovka, Moscow District, Russian Federation;*

[2] *Institute of Physical Chemistry, Polish Academy of Sciences, Warsaw 01-224, Poland.*

Tellurium trioxide, TeO$_3$, is the only example of a trioxide adopting at ambient conditions the VF$_3$-type structure (a distorted variant of the cubic ReO$_3$ structure). Here we present a combined experimental (Raman scattering) and theoretical (DFT modelling) study on the influence of high pressure (exceeding 100 GPa) on the phase stability of this compound. In experiment the ambient-pressure VF$_3$-type structure ($R\bar{3}c$ symmetry) is preserved up to 110 GPa. In contrast, calculations indicate that above 66 GPa the $R\bar{3}c$ structure should transform to a YF$_3$-type polymorph (*Pnma* symmetry) with the coordination number of Te$^{6+}$ increasing from 6 to 8 upon the transition. The lack of this transition in the room-temperature experiment is most probably connected with energetic barriers, in analogy to what is found for compressed WO$_3$. The YF$_3$-type phase is predicted to be stable up to 220 GPa when it should transform to a novel structure of $R\bar{3}$ symmetry and $Z$ = 18. We analyse the influence of pressure on the band gap of TeO$_3$, and discuss the present findings in the context of structural transformations of trioxides and trifluorides adopting an extended structure in the solid state

## Introduction

Tellurium, one of the heaviest metalloids, has recently attracted significant attention due to its role in manufacturing highly efficient photovoltaic panels,[1,2] and investigations of superconductivity in iron-based compounds.[3] In these systems tellurium is found as an anion (Te$^{2-}$), but materials containing this element as a cation are also of great interest, as exemplified by the acousto-optic and nonlinear optoelectronic properties of tellurium dioxide (TeO$_2$).[4,5]

Apart from TeO$_2$, four other oxides of tellurium are known in the solid state: tellurium trioxide (TeO$_3$) and two mixed-valent compounds containing both Te$^{4+}$ and Te$^{6+}$ ions (Te$_4$O$_9$ and Te$_2$O$_5$).[6,7] The phase transitions of TeO$_2$ induced by pressures exceeding 1 GPa ( =10 kbar) were studied intensively both experimentally,[8–13] and by Density Functional Theory (DFT) modelling.[13–16] Compression induces a substantial volume reduction in TeO$_2$ (by about 36 % up to 70 GPa), and an increase in the coordination number (CN) of Te$^{4+}$ from 4 to 9.[12]

In contrast, the high-pressure phase transitions of TeO$_3$, which contains the much smaller Te$^{6+}$ cation, were not studied up to date. One of the probable reasons for such lack of research is connected with the poor availability of the starting material. Tellurium trioxide cannot be prepared by direct oxidation of Te or TeO$_2$, but requires performing the thermal decomposition of orthotelluric acid, Te(OH)$_6$. Moreover, there is some variation in the published recipes for the synthesis.[17–21]

TeO$_3$ is reported to exhibit four different phases: three of them are crystalline (marked I, II, III following the notation introduced in ref. [21]) and one is amorphous (IV). Full structural information is available only for phase I (often referred to as β-TeO$_3$), which adopts a VF$_3$-type structure (Figure 1) exhibiting a coordination number (CN) of Te$^{6+}$ equal to 6.[18,20,21] Phase II is reported to adopt a hexagonal unit cell, while for phase III only information on unindexed powder X-ray diffraction lines is available.[21]

The ambient-pressure VF$_3$-type structure ($R\bar{3}c$ symmetry, $Z = 6$)[20,21] is composed of corner-sharing TeO$_6$ octahedra with all six Te-O contacts equal in length. This geometry can be viewed as a perovskite-type ABX$_3$ structure with vacant A-sites and tilting of the BX$_6$ octahedra. The VF$_3$ polytype can be derived from the non-tilted ReO$_3$ structure ($Pm\bar{3}m$, $Z = 1$)[22] by enforcing tilting of the octahedra ($a^-a^-a^-$ type tilting in Glazer's notation).[23]

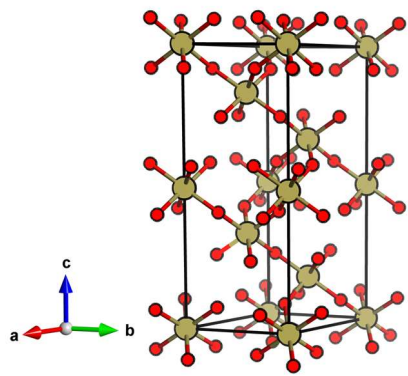

**Figure 1** The ambient-pressure VF$_3$-type structure of TeO$_3$ ($R\bar{3}c$) symmetry.

It's noteworthy to point out that phase I of TeO$_3$ is the only example of a trioxide adopting the VF$_3$-type structure. This polytype is more commonly encountered in trifluorides. In this context, it is of interest to explore the high-pressure phase transformations of TeO$_3$, and compare them with those of trifluorides,[24–26] and trioxides, in particular, WO$_3$,[27–31] and ReO$_3$ itself.[32–35]

Here we present a combined experimental (Raman scattering) and theoretical (DFT modelling) study on the influence of high pressure on the structure stability of TeO$_3$. Our experimental results indicate that the ambient-pressure VF$_3$-type structure ($R\bar{3}c$ symmetry) is preserved up to 110 GPa. In contrast, calculations indicate that above 66 GPa the $R\bar{3}c$ structure should transform to a YF$_3$-type polymorph (*Pnma* symmetry) with a subsequent increase in the CN of Te$^{6+}$ from 6 to 8. This transition is not observed in an experiment most probably due to large energetic barriers, in analogy to what is found for compressed WO$_3$. Another transition from the *Pnma* structure to a rhombohedral phase ($R\bar{3}$, $Z = 18$) with a 10-fold coordination of Te$^{6+}$ is predicted at 220 GPa. We analyse the influence of pressure on the band gap of TeO$_3$ and discuss the present findings in the context of structural transformations of trioxides and trifluorides adopting an extended structure in the solid state (in contrast to SbF$_3$,[36] or AsF$_3$,[37] which form molecular crystals).

## Experimental and computational details

**Sample preparation:** TeO$_3$ was prepared by heating Te(OH)$_6$, purchased from Aldrich, in a thick glass ampoule at a 350 – 450°C temperature range. The best material was obtained by reaction conducted at 450°C for 20 hours. The purity of the sample was verified by powder X-ray diffraction and Raman spectroscopy (see Figure S 1 in Supporting Information).

**High-pressure experiments:** Three high-pressure runs were conducted with the use of a diamond anvil cell (DAC) equipped with diamonds with a bevelled 300 μm tip (bevel angle of 8°). The sample was enclosed by a stainless-steel gasket pre-indented to a thickness of ca. 30 μm. The gasket hole with a radius of 120 μm was laser-drilled. No pressure-transmitting medium was used. The pressure was determined with the use of the ruby fluorescence scale as proposed by Dewaele *et al.*,[38] as well as the shift of the first-order Raman spectra of the diamond anvil as given by Akahama and Kawamura.[39]

**Raman spectroscopy:** The spectra in the first two runs were collected in backscattered geometry using custom designed setup for micro-Raman measurements based on Jobin Yvon THR1000 monochromator equipped with a single grating (with 1200 grooves mm$^{-1}$) giving a resolution of ~1 cm$^{-1}$, notch filters (Keiser Optical Systems) and thermoelectrically cooled (-65 °C) CCD (Horiba Synapse) detection. A He-Ne laser (Melles-Griot) red line (632.8 nm) was used for sample excitation.

The spectra in the third run were acquired with the Alpha300M+ confocal microscope (Witec Gmbh) equipped with a motorized stage. We used a 532 nm laser line delivered to the microscope through a single-mode optical fiber. The laser power at the sample did not exceed 20 mW. The backscattered Raman signal was collected through a 20× long working distance objective, and passed through a multi-mode optical fiber (50 μm core diameter) to a lens based spectrometer (Witec UHTS 300, f/4 aperture, focal length 300 mm) coupled with a back-illuminated Andor iDUS 401 detector thermoelectrically cooled to -60°C. The spectra were collected in the range of Raman shifts from 70 to 1720 cm$^{-1}$ with the use of an 1800 mm grating resulting in a 1.2 cm$^{-1}$ spectral resolution. The acquisition time was 1 s with 30 accumulations. The spectra were post-processed (background subtraction and cosmic-ray removal) with the Project FIVE software (Witec Gmbh). The position of Raman bands was established with the Fityk 1.3.1 software by fitting the observed bands with Pseudo-Voigt profiles.[40] During the experiment we did not observe any Raman bands that could be assigned to the O$_2$ vibron,[41] which excludes decomposition of TeO$_3$ into O$_2$ and lower-valence tellurium oxides.

**DFT calculations:** Periodic DFT calculations of the geometry and enthalpy of various polymorphs of TeO$_3$ utilized the SCAN meta-GGA functional.[42] This functional was found to offer an accurate description of the high-pressure properties for a wide range of compounds.[43–47] We found that it reproduces very well the geometry and the vibration frequencies of the ambient pressure structure of TeO$_3$ (see Table S 1). Thermodynamic stability of various TeO$_3$ polymorphs was judged by comparing their enthalpy ($H$), and thus the calculations formally correspond to $T = 0$ K at which the Gibbs free energy ($G = H - S \cdot T$, where $S$ is the entropy) is equal to $H$.

The projector-augmented-wave (PAW) method was used in the calculations,[48] as implemented in the VASP 5.4 code.[49,50] The cut-off energy of the plane waves was set to 800 eV with a self-consistent-field convergence criterion of 10$^{-8}$ eV. Valence electrons (Te: 5s$^2$, 5p$^4$; O: 2s$^2$, 2p$^4$) were treated explicitly, while standard VASP pseudopotentials were used for the description of core electrons. We verified that using an extended basis set for Te, with the 3d, 4s, and 4p electrons included explicitly, did not alter the obtained results. The $k$-point mesh spacing was set to $2\pi \times 0.03$ Å$^{-1}$. All structures were optimized until the forces acting on the atoms were smaller than 1 meV/Å. Calculations of vibration frequencies (also using SCAN) were conducted with the finite-displacement method with 0.007 Å displacement. We did not apply any scaling of the theoretical vibration frequencies when comparing them with experimental values.

At selected pressures, we additionally calculated the intensity of Γ-point Raman-active vibrational modes using density-functional perturbation theory (DFPT),[51] as implemented in the CASTEP code (academic release version 19.11).[52] Due to the large computational cost of these calculations we employed the Local Density Approximation (LDA). This approach was previously successfully used to model the Raman spectrum of Te(II) and Te(III) oxides,[53,54] as well as the pressure-induced changes in the Raman spectrum of inorganic fluorides.[46,47] In our calculations we employed norm-conserving pseudopotentials and a cut-off energy of 1020 eV. The Raman activity of each vibrational mode ($S_i$) was converted into the the intensity ($I_i$) assuming the following relation:

$$I_i \sim \frac{(\nu_0 - \nu_i)^4}{\nu_i \left(1 - e^{-h\nu_i c / kT}\right)} S_i$$

where $\nu_0$ is the laser frequency, $\nu_i$ is the mode frequency, T is the temperature (taken as equal to 293 K).

We performed evolutionary algorithm searches for lowest-enthalpy structures of TeO$_3$. For this, we used the XtalOpt software (version r12)[55] coupled with periodic DFT calculations utilizing the Perdew-Burke-Ernzerhof (PBE) functional.[56] These searches were conducted at 50, 60 100, 150, and 200 GPa for $Z$ up to 6 (an additional search at 60 GPa with $Z$ = 12 was also conducted) yielding nearly 5 000 crystal structures.

Visualization of all structures was performed with the VESTA software package.[57] For symmetry recognition we used the FINDSYM program.[58] Group theory analysis of the vibrational modes was performed with the use of the Bilbao Crystallographic Server.[59]

## Results and discussion

**High-pressure Raman scattering**

Previous studies showed that at ambient pressure all of the four Raman-active vibrational modes of the $R\bar{3}c$ structure (A$_{1g}$ + 3 × E$_g$) can be detected, with the A$_{1g}$ mode having a considerable higher intensity than the E$_g$ modes.[19,53,54] In our diamond anvil cell (DAC) high-pressure experiments we could follow the frequency change of all of these modes up 35 GPa (Figure 2). Above that pressure, the highest-frequency E$_g$ mode (3E$_g$) could not be detected due to an increase in the background fluorescence. The lowest-frequency E$_g$ mode (1E$_g$) could be followed up to 90 GPa, while the medium-frequency E$_g$ mode (2E$_g$) is shadowed by the A$_{1g}$ band at approximately the same pressure.

Up to the highest pressure reached in this study (110 GPa) the observed shifts in the frequency of the Raman bands are in good accordance with those predicted for the VF$_3$ structure (Figure 3a). However, above 60 GPa we observe a new band developing on the high-frequency side of the A$_{1g}$ mode (Figure 2b). The emergence of this band is most probably a result of non-hydrostatic conditions inside the DAC at large compression which leads, for a part of the sample, to shifting of the frequency of the A$_{1g}$ mode to higher values (see Figure S 2). The changes in the Raman spectrum, observed in all three experimental runs, and at different positions in the DAC, are reversible upon sample decompression with no significant hysteresis.

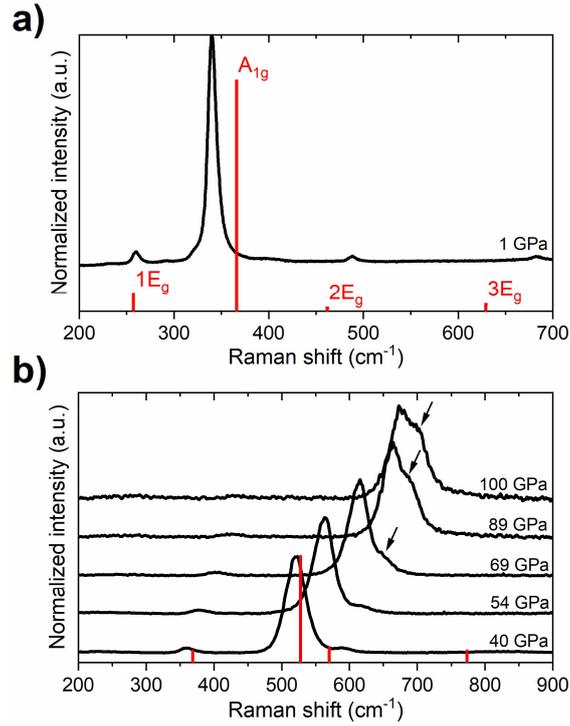

**Figure 2** Raman spectrum (black curves) of the $R\bar{3}c$ polymorph of TeO$_3$ at (a) 1 GPa and higher pressures (b). In (b) the spectra are offset for clarity. Red bars indicate the intensities of Raman bands obtained with LDA simulations performed for the $R\bar{3}c$ structure at 1 and 40 GPa. Arrows mark the position of a new band developing above 60 GPa.

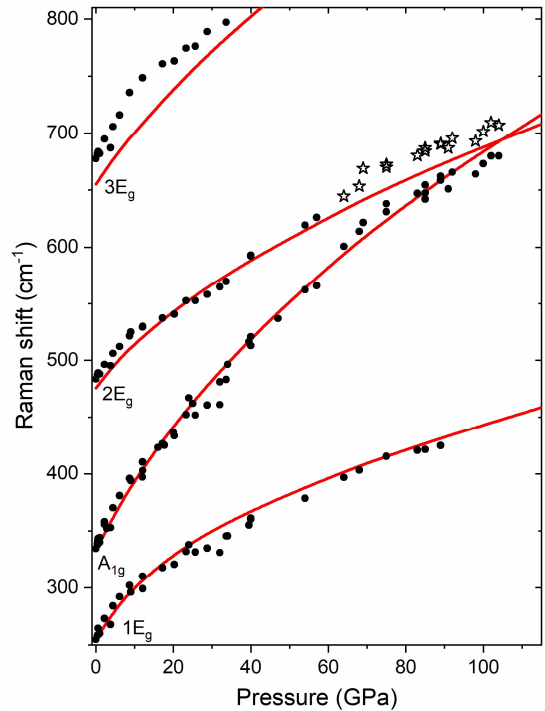

**Figure 3** Pressure dependence of the frequencies of Raman bands of solid TeO$_3$ measured at room-temperature compression (black points). Stars mark the position of the new Raman band which develops above 60 GPa due to non-hydrostatic conditions (see text). Solid red curves mark frequencies of Raman-active bands calculated for the $R\bar{3}c$ structure with the SCAN functional.

**DFT calculations**

To gain more insight in the pressure-induced phase transitions of TeO$_3$, we have conducted evolutionary algorithm searches for the most stable high-pressure polymorphs of this compound (for details see the methods section). These searches identified two low-enthalpy structures (Figure 4a,b): one with *Pnma* symmetry ($Z = 4$), the other with $R\bar{3}$ symmetry ($Z = 18$). The first of these polymorphs is isostructural to YF$_3$ and exhibits 8-fold coordination of Te$^{6+}$ by O$^{2-}$ in the form of a distorted square antiprism. Searches in the Inorganic Crystal Structure Database (FIZ Karlsruhe)[60] indicate that the $R\bar{3}$ structure, exhibiting 10-fold coordination of Te$^{6+}$, has not been previously reported for any AB$_3$-type compound.

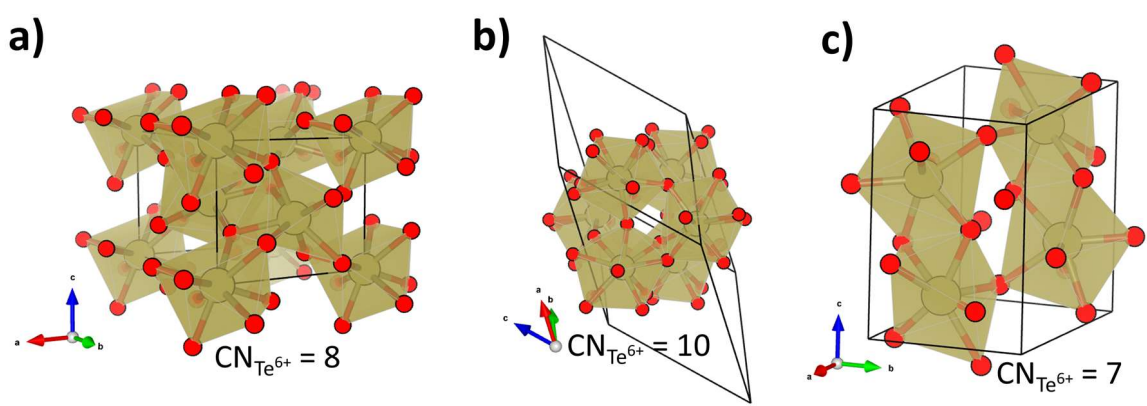

**Figure 4** The high pressure structures of TeO$_3$: YF$_3$-type (*Pnma* symmetry, $Z = 6$) (a), structure of $R\bar{3}$ symmetry ($Z = 18$) (b), HP-WO$_3$-type (*P2$_1$/c*, $Z = 4$) (c). Coordination numbers of Te$^{6+}$ are given for each structure. The $R\bar{3}$ polymorph is shown in the rhombohedral representation.

Analysis of the pressure dependence of the relative enthalpy of various TeO$_3$ polymorphs is given in Figure 5a. Calculations indicate that the VF$_3$-type structure is the most stable one at ambient conditions (≈ 0 GPa) – in accordance with experiment. Upon compression, a first order transition from this structure into the YF$_3$-type polymorph (*Pnma* symmetry) is predicted at 66 GPa. However, the spectral changes observed in the experiment cannot be assigned to this transition, as the predicted Raman spectrum of the *Pnma* structure is very different from the experimental one (Figure S 3). Calculation indicate one more phase transition at 220 GPa from the YF$_3$-type structure to a polymorph of $R\bar{3}$ symmetry. This process leads to an increase in CN from 8 to 10 and a volume reduction of nearly 3 % (Figure 5b).

The lack of the transition between VF$_3$- and YF$_3$-type structures in our room-temperature experiment is most probably due to significant kinetic barriers accompanying this process. These are a result of large changes in the Te coordination environment and molar volume (Figure 5b). To overcome these barriers laser heating of the sample would be required, in analogy to high-pressure experiments conducted for TeO$_2$.[12] However our current experimental setup does not allow us to perform laser heating of samples enclosed in a diamond anvil cell. Moreover leaser heating could lead to thermal decomposition of TeO$_3$ (at ambient conditions this compound decomposes above 400°C).[21]

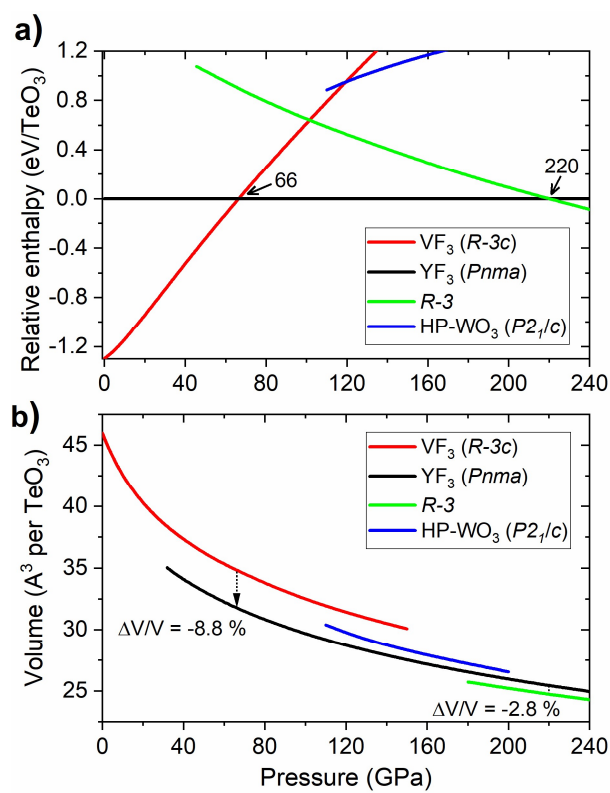

**Figure 5** Pressure dependence of the (a) relative enthalpy (referenced to that of *Pnma*) of the high-pressure polymorphs of TeO$_3$ (SCAN calculations), and (b) their volume. Numbers in (a) indicate the pressures corresponding to the $R\bar{3}c$ → *Pnma* and *Pnma* → $R\bar{3}$ phase transitions. Predicted volume changes at these transitions are given in (b).

The fact that the VF$_3$ structure of TeO$_3$ is metastable up to 110 GPa is further supported by phonon dispersion calculations. In the 0 to 141 GPa pressure range the $R\bar{3}c$ polymorph does not exhibit any vibrations with imaginary frequencies. However, upon compression, a softening of one phonon branch is observed at the F-point of the Brillouin zone, (½, ½, 0) vector, which leads to the appearance of imaginary modes at this point above 141 GPa. This behaviour was also observed at high pressure for the VF$_3$ structure of FeF$_3$.[25]

Both FeF$_3$ and TeO$_3$ exhibit a similar pressure dependence of structural parameters (although in different pressure ranges). For both compounds pressure induces a change in the structural parameters towards an ideal hexagonally close-packing (hcp) packing of O$^{2-}$ anions.[24,25] For example, the *c/a* ratio initially increases upon compression up to a point where it exceeds that predicted for a perfect hcp packing (Figure 6). After reaching a maximum (found at 20 GPa for FeF$_3$, and 80 GPa for TeO$_3$ – see Figure 6) this ratio decreases upon compression. Interestingly in both FeF$_3$ and TeO$_3$ the F-point phonon instability appears at the point when this ratio becomes again smaller than the value predicted for a perfect hcp packing.

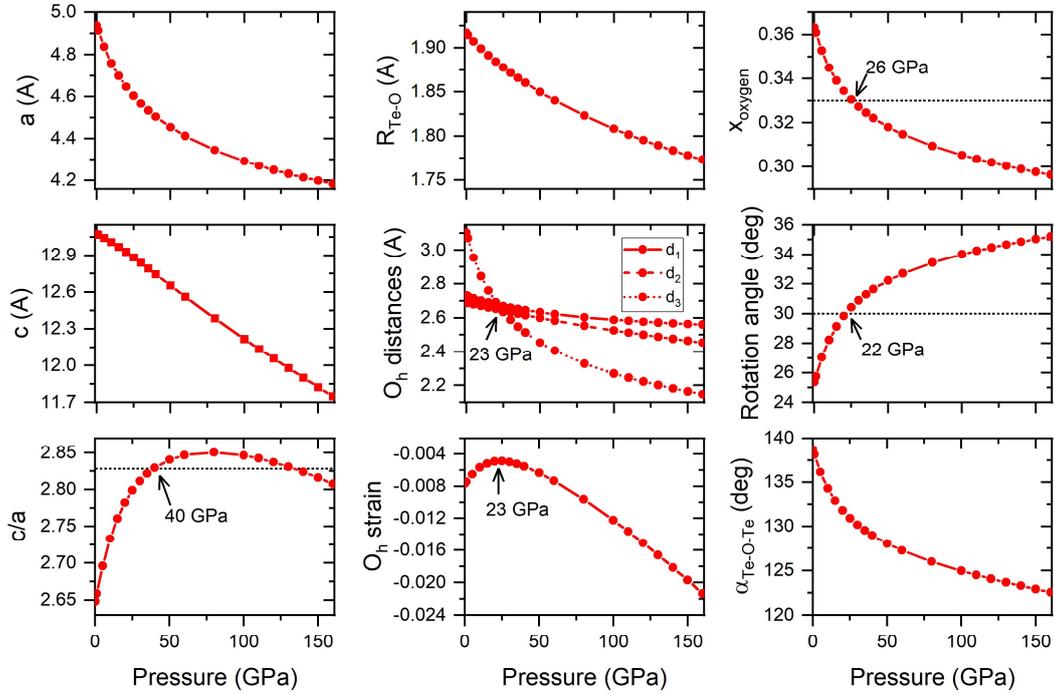

**Figure 6** Calculated evolution of the structural parameters of the $R\bar{3}c$ structure upon compression. The $d_1$ and $d_2$ distances are O-O distances within the TeO$_6$ octahedra ($d_1$ within the ***ab*** plane, $d_2$ out of plane), while $d_3$ is the O-O distance between the octahedra located in the same ***ab*** plane. The octahedral ($O_h$) strain is calculated as equal to $(d_2 - d_1)/(d_2 + d_1)$, negative values indicate compression of the TeO$_6$ octahedron along the ***c*** cell vector. The rotation angle is calculated with respect to the <111> direction of the ideal cubic ReO$_3$ cell; $\alpha_{Te-O-Te}$ is the angle of the Te-O-Te bridge. Dotted horizontal lines indicate $c/a$, $x_o$, and rotation angle values expected for a perfect hexagonal close-packing packing of O$^{2-}$ anions.

Distorting the $R\bar{3}c$ structure along the eigenvector of the imaginary mode, and subsequent geometry optimization, leads to a structure with four TeO$_3$ units per unit cell and $P2_1/c$ symmetry (this space group is the highest-symmetry sub-group of $R\bar{3}c$ along the (½, ½, 0) modulation vector). This structure, shown in Figure 4c, is a distorted variant of the high-pressure phase of WO$_3$ (HP-WO$_3$),[28] and exhibits 7-fold coordination of Te$^{6+}$ (increasing to 8-fold upon compression). The $R\bar{3}c \rightarrow P2_1/c$ transition is of first order ($\Delta V/V = -5.8$ % at 141 GPa). However, given the fact that it is connected with a phonon instability, it might be characterized by a small energetic barrier. It, therefore, could be observed during room-temperature compression.

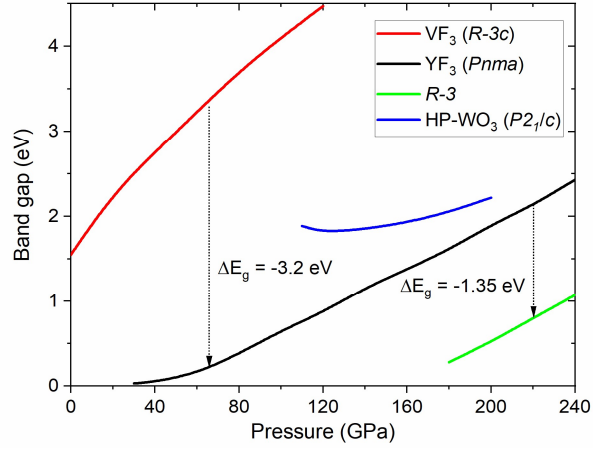

**Figure 7** Dependence of the electronic band gap upon compression for the high-pressure structures of TeO$_3$. Predicted changes in E$_g$ at the $R\bar{3}c$ → *Pnma* (*p* = 60 GPa) and Pnma → $R\bar{3}$ (*p* = 220 GPa) transitions are given.

SCAN calculation of the pressure dependence of the electronic band gap ($E_g$) for the high-pressure structures of TeO$_3$ is shown in Figure 7. The $E_g$ value obtained at ambient conditions for the VF$_3$ structure (1.55 eV) is underestimated compared with the experimental value (3.25 eV), but larger than that obtained in calculations utilizing the PBE functional (1.21 eV).[61,62] Compression induces an increase of the band gap in all TeO$_3$ polymorphs, especially for the VF$_3$-type structure exhibiting octahedral coordination of Te$^{6+}$. This trend is similar to what is found for ionic difluorides (MF$_2$, M = Be, Mg, Ca),[63] and in contrast to TeO$_2$ where much smaller variations of $E_g$ are seen.[15] The band gap is predicted to decrease abruptly upon the $R\bar{3}c$ → *Pnma* (*p* = 60 GPa) and the *Pnma* → $R\bar{3}$ (*p* = 220 GPa) phase transition.

## Conclusions

High-pressure Raman measurements indicate that the ambient-pressure VF$_3$-type structure ($R\bar{3}c$ symmetry) persists up to 110 GPa. Calculations confirm the dynamic stability of this polymorph up to a pressure of 141 GPa. At larger compression a phonon instability should lead to a transition into a distorted variant of the high-pressure phase of WO$_3$ ($P2_1/c$ symmetry) with a subsequent increase of the Te$^{6+}$ coordination number from 6 to 7. We argue that a similar transition could be observed for FeF$_3$, which exhibits the same type of instability.[25] The HP-WO$_3$ phase can also be a candidate for a post-VF$_3$ phase of ReO$_3$.[32]

Calculations indicate that the persistence of the VF$_3$-type structure is a kinetic effect as both this polymorph and the HP-WO$_3$ structure are thermodynamically less stable than a YF$_3$-type polymorph (*Pnma* space group), which exhibits 8-fold coordination of Te$^{6+}$. The lack of the VF$_3$ → YF$_3$ phase transition in room-temperature compression experiments is most probably a result of the large energetic barriers associated with it. It's noteworthy to mention that a similar situation is found for WO$_3$. For this compound calculations indicate that a YF$_3$ structure is the ground state structure above 20 GPa.[31] However, it has not been observed in experiments conducted up to 40 GPa.[29]

We predict that above 220 GPa TeO$_3$ should enter a post-YF$_3$ phase with $R\bar{3}$ symmetry (Z = 18) and 10-fold coordination of Te$^{6+}$. We do not find evidence for a possible transition into the 12-fold coordinated LaF$_3$-type structure ($P\bar{3}c1$) adopted by rare earth trifluorides, nor to the 9-fold coordinated *Cmcm* phase proposed for WO$_3$.[31] Finally, we show that high pressure and compression-induced phase transitions lead to large variations of the electronic band gap of TeO$_3$. The possibility of pressure-tuning of the band structure of TeO$_3$ polymorphs is also of interest in the context of non-linear properties of this material.[61]

**Acknowledgments**: This research was carried out with the support of the Interdisciplinary Centre for Mathematical and Computational Modelling (ICM), University of Warsaw, under grants no. GB74-8 and GB80-11.

**Notes**: The authors declare no competing financial interest.

**Supporting Information Available:** Comparison of the ambient-pressure geometry and the frequencies of the Raman-active vibrational modes between experiment and SCAN calculations; ambient-pressure powder X-ray diffraction pattern of TeO$_3$; comparison of the high-pressure Raman spectrum of TeO$_3$ with that calculated for the YF$_3$-type and VF$_3$-type structures; computed structural parameters of the high-pressure polymorphs of TeO$_3$.

**Supporting Information**

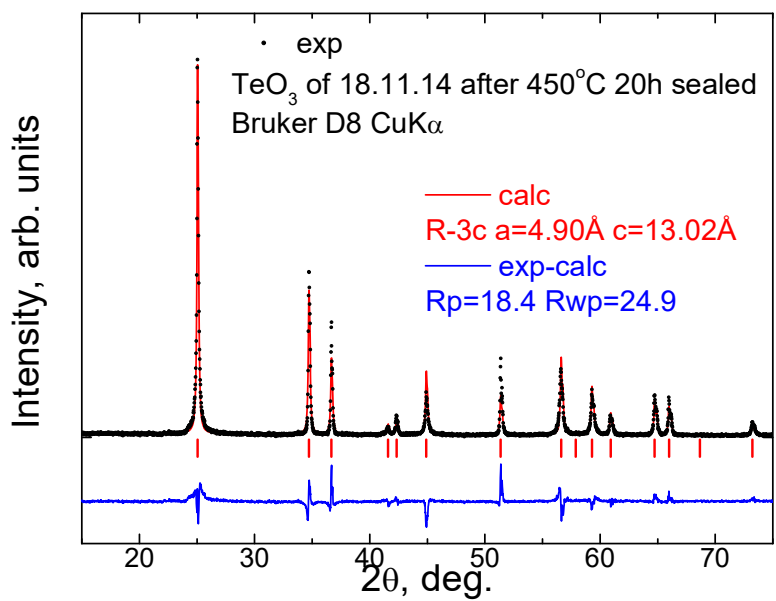

**Figure S 1** Powder X-ray diffraction pattern of freshly synthesized TeO₃ (black points) obtained using Bruker D8 diffractometer and CuK$_\alpha$ radiation. The Rietveld fit assuming the $R\bar{3}c$ structure is shown with a red curve. Blue curve represents the difference between the fit and experimental points.

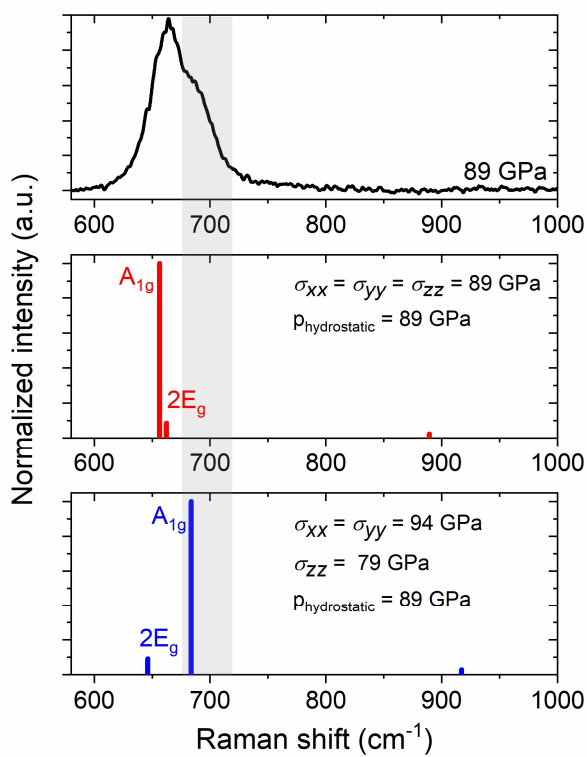

**Figure S 2** Raman spectrum (black curve) of TeO₃ at 89 GPa compared with the Raman band intensities simulated with LDA for the $R\bar{3}c$ structure at a hydrostatic pressure of 89 GPa (red bars), and at non-hydrostatic conditions (blue bars).

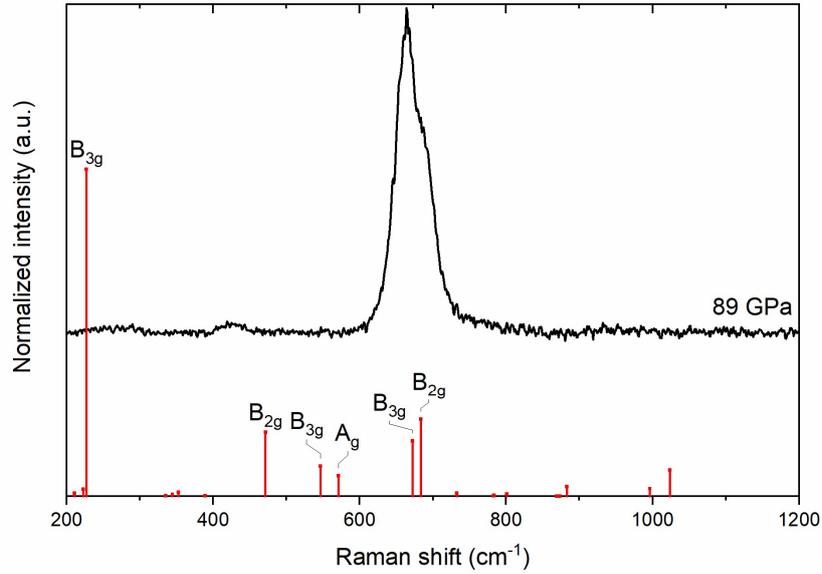

**Figure S 3** Comparison of the Raman spectrum of TeO$_3$ at 89 GPa (black curves) with the Raman intensities simulated at that pressure for the YF$_3$-type (*Pnma* symmetry) structure (red bars).

**Table S 1** Comparison of the experimental geometry (from this work and ref. [20]) and the frequencies of the Raman-active vibrational modes (this work and ref. [19]) of the ambient pressure polymorph of TeO$_3$ (space group $R\bar{3}c$) with data obtained from calculations utilizing the SCAN and LDA functionals. Cell vectors and the Te-O bond length are given in Å, volume in Å$^3$, while frequencies in cm$^{-1}$. Percentage differences between our experimental data and calculations are given in parenthesis.

|  | Exp. | Exp. (this work) | SCAN (this work) | LDA (this work) |
|---|---|---|---|---|
| **a** | 4.901 | 4.900 | 4.933 (+0.7 %) | 4.959 (+1.2 %) |
| **c** | 13.030 | 13.020 | 13.074 (+0.4 %) | 13.314 (+2.3 %) |
| **V/Z** | 45.17 | 45.12 | 45.91 (+1.8 %) | 47.25 (+4.7 %) |
| **R$_{Te-O}$** | 1.911 | 1.910 | 1.917 (+0.4 %) | 1.975 (+3.4 %) |
| **1E$_g$** | 258 | 255 | 255 (–0.0 %) | 251 (–1.6 %) |
| **A$_g$** | 336 | 334 | 333 (–0.3 %) | 360 (+7.8 %) |
| **2E$_g$** | 485 | 484 | 476 (–1.7 %) | 457 (–5.6 %) |
| **3E$_g$** | 666 | 677 | 655 (–3.2 %) | 624 (–7.8 %) |

**Table S 2** Computed (SCAN calculations) structural parameters of the high-pressure polymorphs of TeO$_3$.

| | a | b | c | α | β | γ | V | Fractional coordinates |
|---|---|---|---|---|---|---|---|---|
| **VF$_3$ ($R\bar{3}c$)** <br> **p = 60 GPa** | 4.412 | 4.412 | 12.561 | 90 | 90 | 120 | 211.78 | O1 *18e* (0.685 0.0 0.25) <br> Te1 *6b* (0.0 0.0 0.0) |
| **HP-WO$_3$ ($P2_1/c$)** <br> **p = 110 GPa** | 4.809 | 4.343 | 5.857 | 90 | 96.15 | 90 | 121.61 | O1 *4e* (0.052 0.663 0.846) <br> O2 *4e* (0.615 0.0146 0.347) <br> O3 *4e* (0.724 0.509 0.455) <br> Te1 *4e* (0.743 0.683 0.161) |
| **YF$_3$ (*Pnma*)** <br> **p = 100 GPa** | 4.888 | 5.960 | 4.074 | 90 | 90 | 90 | 118.69 | O1 *8d* 0.670 0.065 0.153) <br> O2 *4c* (0.525 0.25 0.623) <br> Te1 *4c* (0.871 0.250.465) |
| **$R\bar{3}$** <br> **p = 100 GPa** | 5.851 | 5.851 | 17.358 | 90 | 90 | 120 | 514.77 | O1 *18f* (0.352 0.376 0.303) <br> O2 *18f* (0.745 -0.027 0.186) <br> O3 *6c* (0.0 0.0 0.715) <br> O4 *6c* (0.0 0.0 0.437) <br> O5 *6c* (0.0 0.0 -0.074) <br> Te1 *18f* (0.003 0.326 0.251) |